\documentclass[final,5p,times,twocolumn]{elsarticle}

\usepackage{bm} 
\usepackage{graphicx}
\usepackage{dcolumn}
\usepackage{gensymb}
\usepackage{hyperref}
\hypersetup{
    colorlinks=true,
    linkcolor=blue,
    filecolor=blue,      
    urlcolor=blue,
    citecolor= blue,}

\usepackage{amsmath}
\usepackage{amssymb}
\usepackage{booktabs}
\usepackage{arydshln}

\journal{Physics Letters B}

\begin{document}

\title{Isomeric yield ratios and mass spectrometry of Y and Nb isotopes in the neutron-rich N=60 region: the unusual case of $^{98}$Y}
\author[Uppsala]{Simone Cannarozzo}
\ead{simone.cannarozzo@physics.uu.se}
\author[Uppsala]{Stephan Pomp}
\ead{stephan.pomp@physics.uu.se}
\author[Jyv]{Anu Kankainen}
\author[Jyv]{Iain Moore}
\author[Jyv]{Marek Stryjczyk}
\author[Uppsala]{Ali Al-Adili}
\author[Uppsala]{Andreas Solders}
\author[Jyv]{Ville Virtanen}
\author[Jyv]{Tommi Eronen}
\author[Uppsala]{Zhihao Gao}
\author[Jyv]{Zhuang Ge}
\author[Jyv]{Arthur Jaries}
\author[Uppsala]{Mattias Lantz}
\author[Jyv]{Maxime Mougeot}
\author[Jyv]{Heikki Penttil{\"a}}
\author[Jyv]{Andrea Raggio}
\author[Jyv]{Jouni Ruotsalainen}
\affiliation[Uppsala]{organization={Department of Physics and Astronomy, Uppsala University},
            addressline={Box 516}, 
            city={Uppsala},
            postcode={751 20}, 
            country={Sweden}}
\affiliation[Jyv]{organization={Department of Physics, Accelerator Laboratory, University of Jyvaskyla},
            addressline={ P.O. Box 35(YFL)}, 
            city={Jyvaskyla},
            postcode={40014}, 
            country={Finland}}

\begin{abstract}
The isomeric yield ratio (IYR) of fission products is an observable that carries relevant information about the fragments emerging from the scission of a fissioning nucleus.
We report on IYR of $^{96,98,100}$Y and $^{100,102}$Nb, together with the previously reported values for $^{97}$Y and $^{99}$Nb, produced in the 28 MeV $\alpha$-induced fission of $^{232}$Th at the Ion Guide Isotope Separation On-Line (IGISOL) facility of the University of Jyv{\"a}skyl{\"a}. 
We measured the IYR using two different techniques, the phase-imaging ion-cyclotron-resonance (PI-ICR) and the multiple-reflection time-of-flight mass spectrometry (MR-TOF-MS) methods.
Moreover, we measured the masses of the long-lived states in $^{98,100}$Y and $^{100,102}$Nb populated via in-trap $\beta$-decay of their precursors. 
Since the $\beta$-decay selectively populates states with a favourable spin-parity, we could identify the measured state and show that the ground state is the low-spin state in the cases of $^{98}$Y and $^{100}$Nb, while it is the high-spin state in the cases of $^{100}$Y and $^{102}$Nb. This measurement confirms the spin-parity assignments of all the nuclei as they are reported in the NUBASE2020 evaluations, disagreeing with the assignment for $^{100}$Y reported in the ENSDF evaluation.

Making also use of previously reported data, we observe an anomalously low IYR for the $N=59$ isotope $^{98}$Y as compared to other yttrium or neighboring niobium isotopes. 
This behavior is very rare across the nuclear chart and is posited to be connected to the characteristic shape coexistence of $^{98}$Y, and to the change in the charge radii of the ground and excited states in the $N=58-60$ region.

\end{abstract}

\maketitle

\section{Introduction}

In binary fission, two highly excited fission fragments (FFs) are produced. These fragments de-excite via prompt-neutron and $\gamma$-ray emission, resulting in what are commonly referred to as fission products (FPs) \cite{andreyevNuclearFissionReview2018,Schmidt2018,SCHUNCK2022103963}.
The de-excitation of the FPs can populate either the ground state or a long-lived excited state, which may further decay via internal transition (IT) or $\beta$-decay.\label{sec:intro}

The relative population of isomeric states, the so-called isomeric yield ratio (IYR), is influenced, among other factors, by the angular momentum of the fission fragment (FFs) emerging from scission, and by the spins of the isomeric states \cite{Vandenbosch60,Huizenga60,Al-Adili2019,GaoAM}.
Here we define the IYR as:

\begin{equation}
  \text{IYR} = \frac{\text{Y}_{\text{hs}}}{\text{Y}_{\text{hs}}+\text{Y}_{\text{ls}}}, 
  \label{eq:IYR_def}
\end{equation}

where Y$_{\text{hs}}$ and Y$_{\text{ls}}$ represent the FP yields of the high-spin and low-spin states, for an isomeric pair.

The FFs emerging from scission typically carry a significant amount of angular momentum, regardless of the fissioning system \cite{Wilhelmy,wilsonAngularMomentumGeneration2021}. As a result, during de-excitation, FPs preferentially populate higher spin states, \textit{i.e.} the IYR values are typically larger than 0.5 \cite{RakAM_piicr,RakAM_tof,Sears2021118,Gao2023_iyrs,cannarozzo2024disentanglinginfluenceexcitationenergy}.

In this work we present the IYR measurement of $^{96,98,100}$Y and $^{100,102}$Nb produced by the 28 MeV $\alpha$-particle induced fission of $^{232}$Th at the Ion Guide Isotope Separation On-Line (IGISOL) facility (Table \ref{tab:IYR}). We also utilize the IYRs of $^{97}$Y and $^{99}$Nb obtained in the same experimental campaing but presented in Ref. \cite{cannarozzo2024disentanglinginfluenceexcitationenergy}.
\begin{table*}[h]
    \centering
\caption{\label{tab:IYR}
%
%
Measured IYR of Y and Nb isotopes from $^{232}$Th($\alpha$,f) and relevant nuclear data. All IYRs were measured using the PI-ICR technique, except for $^{96}$Y, which was measured using the MR-TOF-MS. The values of half-life (t$_{1/2}$), spin (I$^{\pi}$) and excitation energy (E$_{x}$) are taken from the NUBASE2020 evaluation \cite{NUBASE2020}. The excitation energy of $^{96}$Y was taken from Ref. \cite{Iskra2022}. The IYRs of nuclei marked with (*) were first reported in Ref. \cite{cannarozzo2024disentanglinginfluenceexcitationenergy}.}
\begin{tabular}{|l|cc|ccc|c|}
\hline

\multicolumn{1}{|c|}{} & \multicolumn{2}{c|}{Ground state} & \multicolumn{3}{c|}{Excited state}&\multicolumn{1}{c|}{}\\
Nuclide & t$_{1/2}$ (s) & I$^{\pi}$ &  t$_{1/2}$ (s)  & I$^{\pi}$ & E$_{x}$ (keV) & IYR \\
\hline
$^{96}$Y & $5.34(5)$ & $0^-$ & $9.6(2)$ & $8^+$& 1540.5(4) & $0.58(2)$ \\
$^{97}$Y (*) & $3.75(3)$ &$ 1/2^-$& $1.17(3)$ & $9/2^+$& $667.52(23)$& 0.800(5)  \\
$^{98}$Y & $0.548(2)$ & $0^-$  & $2.32(8)$ & $(6^+,7^+)$ & $465.7(7)$ & 0.129(4) \\
$^{100}$Y & $0.94(3)$ & $4^+$ & $0.727(6)$ & $1^+ $ & $144(16) $ & 0.727(9) \\
$^{99}$Nb (*) & $15.0(2)$ & $9/2^+$ & $150(12) $ & $1/2^-$ & $365.27(8)$ & 0.732(9) \\
$^{100}$Nb & $1.5(2)$ & $1^+$  & $2.99(11)$ & $(5^+)$& $313(8)$ & 0.80(1) \\
$^{102}$Nb & $4.3(4)$ & $(4^+)$ & $1.31(16)$ & $(1^+)$&  $94(7)$  &  0.74(1)  \\
\hline
\end{tabular}

\end{table*}

The nuclei under analysis belong to a unique region in the nuclear chart, that has been intensively studied in the past years \cite{Garrett2022,Togashi2016}. 
Nuclei with proton number Z between 37 and 41 (Rb, Sr, Y, Zr, and Nb) show a sudden onset of the ground state deformation when crossing the neutron number $N=60$.

Laser spectroscopy data suggest that the ground state shape of yttrium isotopes is near-spherical at the neutron shell closure $N=50$, but becomes increasingly more oblate when moving toward more neutron-rich isotopes. 
At $N=60$, the ground-state shape shifts into a strongly deformed prolate configuration. 
The $N=59$ isotope $^{98}$Y represents the transitional nucleus \cite{Cheal,Urban2017} where the two shapes coexist.
Although the ground state of $^{98}$Y does not allow a direct measurement of the quadrupole moment, a probe of the corresponding deformation, it is assumed that its shape follows the trend of the lighter isotopes. At the same time, $^{98m}$Y exhibits a similar prolate shape to that of $^{99}$Y.

\begin{figure}[]
    \centering
    \includegraphics[scale = 0.87]{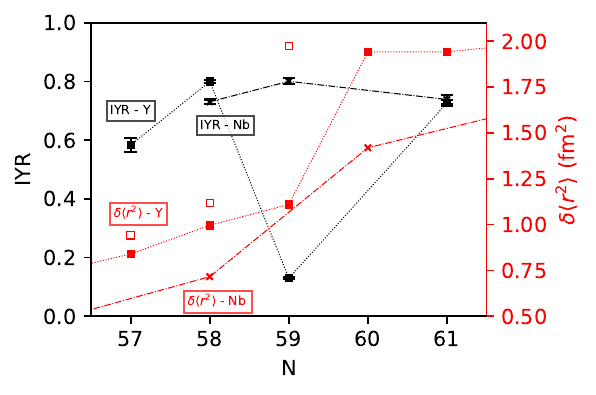}
    \caption{In black, IYR of Y ($Z=39$) and Nb ($Z=41$) isotopes produced by the 28 MeV $\alpha$-particle induced fission for $^{233}$Th as a function of neutron number N. In red, the changes in the mean-square charge radii $\delta\langle r^2 \rangle$ (data from Ref. \cite{ANGELI201369_rms}) compared to $^{89}$Y and $^{93}$Nb for  yttrium and niobium isotopes, respectively. Hollow markers correspond to $\delta\langle r^2 \rangle$ of the isomeric states (data taken from Ref. \cite{Cheal}). }
    \label{fig:IYR}
\end{figure}

Additionally, the mean-squared charge radius of $^{98m}$Y is closer to that of $^{99}$Y than that of $^{98}$Y.
Experimental $\delta\langle r^2 \rangle^{A,A'}$ data \cite{ANGELI201369_rms,Cheal} show that the large change between the two isomeric states $\delta\langle r^2 \rangle^{98,98m}$ = 0.86 fm$^2$ is even greater than the one between the two ground states $\delta\langle r^2 \rangle^{98,99}$ = 0.82 fm$^2$ (See Fig.~\ref{fig:IYR}). 

\section{Experimental methods}



The fission products were produced using a 10 mg/cm$^2$ $^{232}$Th target irradiated with 32 MeV $\alpha$-particles from the K-130 cyclotron, resulting in an estimated average energy of 28 MeV within the target. The positively charged FPs were then extracted by 30 kV electrostatic potential, and transported to a 55$\degree$ dipole magnet for isobar selection. Finally, the continuous beam of fission products was collected in a radiofrequency quadrupole cooler-buncher (RFQ-CB) \cite{Nieminen,VIRTANEN2025170186}, and delivered as short ion bunches to either the Multi-Reflection Time-of-Flight Mass Spectrometer (MR-TOF-MS) \cite{Virtanen_tbs} or to the JYFLTRAP double Penning trap mass spectrometer. Further experimental details are provided in the following subsections \cite{Eronen20121}. Further experimental details are given in the following subsections.

In total, we determined the IYR of 5 FPs. Of these, the MR-TOF-MS was applied for $^{96}$Y (Sec. \ref{sec:mrtof}) and the PI-ICR technique in the JYFLTRAP for $^{98,100}$Y and $^{99,100}$Nb (Sec. \ref{sec:piicr}). 

In addition, we performed dedicated mass measurements employing the so-called in-trap decay method for $^{98,100}$Y and $^{100,102}$Nb to determine the ordering of the isomeric states, \textit{i.e.} identifying which is the high-spin and which is the low-spin state.
For $^{100}$Y, the ENSDF \cite{SINGH20211}
and Nubase 2020 evaluations (based on Ref. \cite{Baczynska_2010}) do not agree as they report the opposite state ordering.  
In the case of $^{102}$Nb, the ENSDF evaluation \cite{DEFRENNE20091745} explicitly calls for a dedicated measurement for the state ordering of the two isomers.

\subsection{IYR measurement with the MR-TOF-MS for $^{96}$Y}
\label{sec:mrtof}
While MR-TOF-MS are well established devices for high-precision nuclear mass measurements, and for the search for new isomers \cite{DICKEL2015172,Dickel2024,Wolf2013123_isoltrap_mrtof,SCHURY201439_riken_mrtof,Rosenbusch2023_riken_mrtof,Kwiatkowski2024_titan_mrtof,Chauveau2016211_ganil_mrtof,Chauveau2016211_ganil_mrtof,Schultz2016251_nd_mrtof}, their use for IYR measurements is less common \cite{Ayet2019}.
The method is based on the fact that ions with the same energy but different mass-over-charge ratios ($m/q$) disperse in time-of-flight (TOF) according to ${t=a\sqrt{m/q} +b}$, where the parameter $a$ depends on the ion path and the parameter $b$ is a constant. 


During this measurement, the ions were trapped and reflected between electrostatic mirrors in the MR-TOF-MS, so that the time-of-flight could be increased to separate relative mass-over-chrage differences around $10^{-5}$ within 10-30~ms TOF. This resolution was sufficient to resolve mass differences above roughly 1~MeV/$c^2$ at 96~u, meaning that isobars, as well as isomeric states with excitation energies of this order or higher, could be separated.

\begin{figure}[h]
    \centering
    \includegraphics[scale = 0.9
]{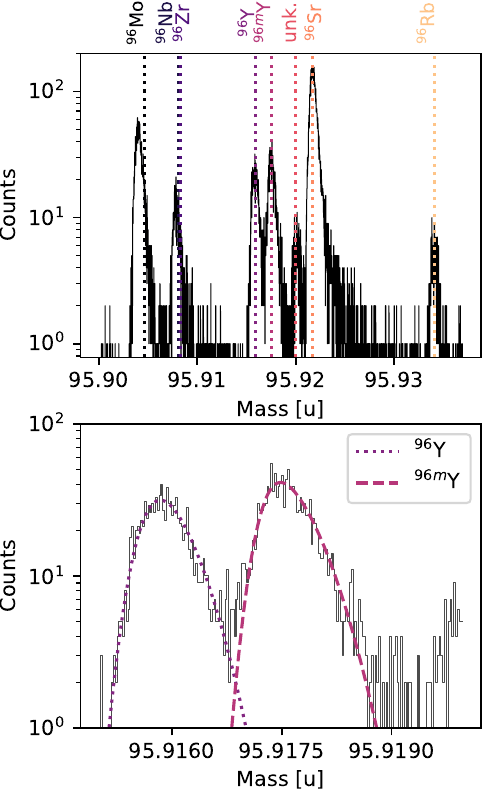}
    \caption{MR-TOF-MS mass spectrum of fission products with mass $A = 96$ produced by the $^{232}$Th($\alpha$,f) fission reaction. The top plot shows the complete spectrum, while the bottom plot presents a zoom in of the nuclei of interest, \textit{i.e.} the isomeric states of $^{96}$Y. The time of flight of $^{96}$Y was 28.254 ms after 864 revolutions. 
    }
    \label{fig:mrtof}
\end{figure}

The MR-TOF-MS mass spectrum for A=96 is shown in Figure~\ref{fig:mrtof}. For the IYR measurement of $^{96}$Y, 864 revolutions in the trap were required to resolve the isomer from the ground state. 
The peaks corresponding to the two isomeric states of $^{96}$Y are fitted using a skewed gaussian distribution. 
The fit is used to calculate the number of counts in the peaks and the resulting yield ratio as $IYR_{exp} = {C_{hs}}/({C_{hs}+C_{ls}})$. 
The measurement was performed three times, and the reported IYR is the weighted average of the three single values further corrected for the decay losses during the transport (more details in Refs. \cite{CanPHD}). 

\subsection{IYR measurement with the PI-ICR technique at JYFLTRAP}
\label{sec:piicr}
The phase-imaging ion cyclotron resonance (PI-ICR) technique \cite{EliseevPiicr, Nesterenko2018,nesterenkoStudyRadialMotion2021} was initially applied to high-precision mass spectrometry but is also a well-established method to measure the IYR of FPs, as shown in Refs. \cite{RakAM_piicr,Gao2023_machine,Gao2023_iyrs,cannarozzo2024disentanglinginfluenceexcitationenergy}.
The measurements included in this work are part of an experimental campaign already presented in detail in Refs. \cite{cannarozzo2024disentanglinginfluenceexcitationenergy,CannarozzoConference}.

In the PI-ICR technique, Penning traps were used to spatially separate ions. 
Ions with different mass accumulated a different phase within a certain applied phase accumulation time $t_{acc}$. 
Thus, even low-lying excited states ($E_{x} <$  100 keV) could be resolved from the ground state. 

\begin{figure}[h]
    \centering
    \includegraphics[scale = 0.75]{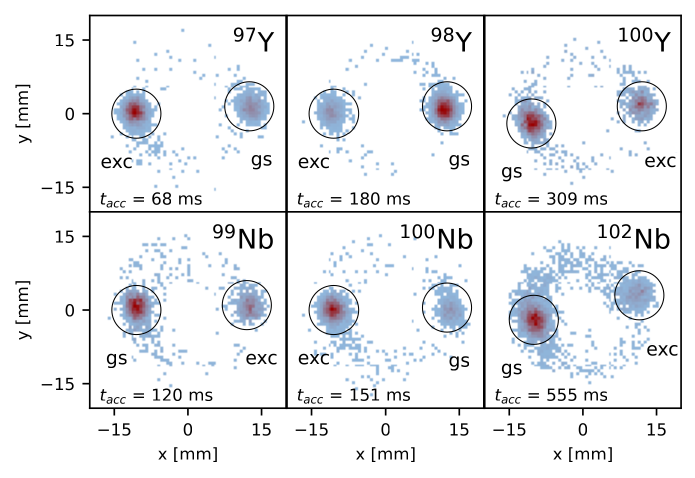}
    \caption{PI-ICR images for the studied Y and Nb isotopes. In each figure, the ground state (gs) and excited state (exc) have been indicated as well as the phase accumulation time.}
    \label{fig:piicr}
\end{figure}

The isobaric ion bunch from the RFQ-CB contained not only the ions of interest but also other isobars. To remove these, the mass-selective buffer-gas cooling method \cite{SAVARD1991247} was applied in the first trap of JYFLTRAP. 
In the second trap, the ground and the excited states were spatially separated. 
Finally, the ions were extracted from the second trap and detected by a position sensitive microcannel plate detector (2D MCP). 
The phase accumulation time was selected to resolve the states at distinct but consistent positions on the detector.
Figure~\ref{fig:piicr} shows the detected images for the studied nuclides. 

The number of counts in each spot was determined by counting all ions detected within 3$\sigma$ of the spots' centers. The IYR is finally evaluated based on the number of counts in the spots, with corrections applied for factors such as spots tails, detector spatial non-uniformity, and decay losses during transport. (more details in Refs.  \cite{cannarozzo2024disentanglinginfluenceexcitationenergy,CannarozzoConference,CanLic}).


\begin{figure*}[h]
    \centering
    \includegraphics[scale = .8]{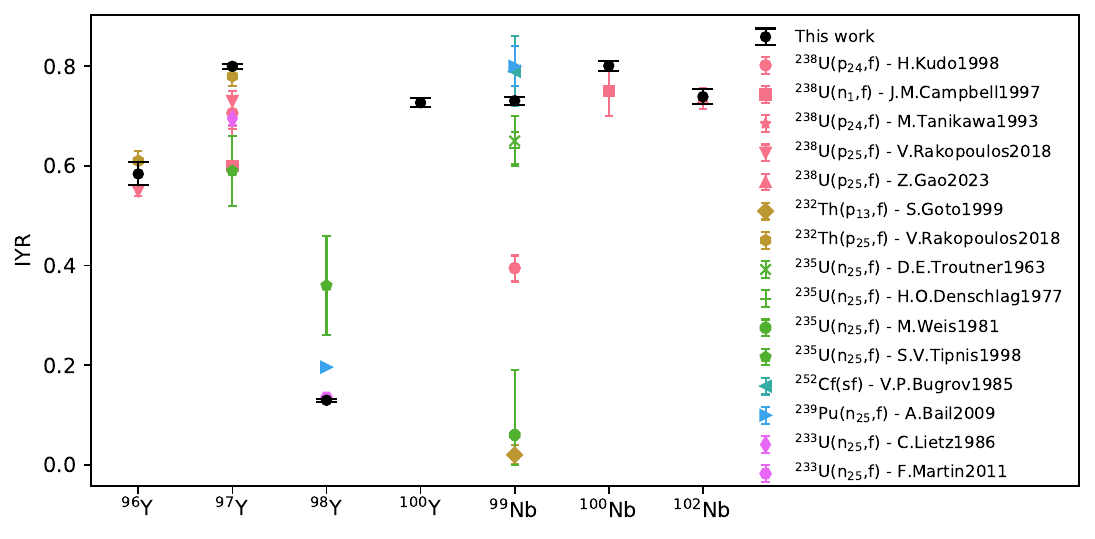}
    \caption{IYR of Y and Nb isotopic series compared to bibliographic data from different fissioning systems retrieved from the nuclear data library EXFOR \cite{EXFORExperimentalNuclear}. The reason for the disagreement on the IYR of $^{99}$Nb is not clearly identifiable, but might be related to using an outdated decay scheme in the $\gamma$-spectroscopy measurements. Data taken from Refs. 
    \cite{Kudo1998KU01,Campbell1997CAZW,Tanikawa199353,RakAM_tof,Gao2023_iyrs,Goto1999GO06,Troutner1963TR01,Denschlag1977_exfor20878.1,Weis1981WE20,Tipnis,Bugrov1985_exfor40895.1,Bail,Lietz1986_exfor22017.1,Martin}
    }
    \label{fig:IYR_bib}
\end{figure*}

\subsection{Mass spectrometry of  $^{98,100}$Y and $^{100,102}$Nb with selective in-trap beta decay}
A dedicated measurement was performed to identify the low- and high-spin states of $^{98,100}$Y and $^{100,102}$Nb, and to verify their masses using the time-of-flight ion cyclotron resonance (TOF-ICR) \cite{Eronen20121, KONIGTOF} technique with the JYFLTRAP Penning trap.
In this measurement, the FPs were produced by the proton-induced fission of $^{238}$U  at an average energy of 24 MeV.

The idea behind the technique is to trap precursor nuclei in the first Penning trap, which mainly decay by $\beta$-emission into a specific isomeric state due to spin selection rules \cite{HukkanenInTrap,HukkanenInTrap2}.
By measuring the mass of the decay product in the second trap, and comparing it to known values, the spin can be linked to a specific mass, thereby unambiguously determining the ordering of the isomeric states. In this measurement, the TOF-ICR technique was applied as at the time of the measurement, the PI-ICR method was not available.

In the following, $^{98}$Y will be used as an example to illustrate the technique. A more detailed description and the results of the in-trap decay measurements will be reported in a separate publication \cite{Ruotsalainen_tbs}.

The precursor of $^{98}$Y, $^{98}$Sr ($t_{1/2}=653(20)$ ms) was allowed to $\beta$-decay in the first trap. It has a spin 0$^+$ and therefore dominantly populates the low-spin state of $^{98}$Y \cite{SINGH20211}. The produced doubly charged $^{98}$Y$^{2+}$ ions were centered in the first trap using the buffer-gas cooling technique \cite{SAVARD1991247}. The ions were sent to the second trap where their mass was determined using the TOF-ICR technique \cite{Eronen20121,KONIGTOF}.   

The measured mass for the low-spin state of $^{98}$Y matched the literature for the ground state within the uncertainties. Thus, we can unambiguously assign the low-spin state as the ground state and the high-spin state as the excited one. 

\section{Results and discussion}

The measured IYRs are reported in Table~\ref{tab:IYR} and Figure~~\ref{fig:IYR}. 
Additionally, through the in-trap decay and TOF-ICR we found that for $^{98}$Y and $^{100}$Nb the ground state is the low-spin state. 
In the cases of $^{100}$Y and $^{102}$Nb, the state ordering is the opposite.
This confirms the state ordering of the NUBASE2020 evaluations of 
these four nuclei as given in \cite{NUBASE2020}, 
and reverses the ENSDF one in the case of $^{100}$Y \cite{SINGH20211}, 
clarifying a disagreement between the ENSDF and NUBASE2020 evaluations.
%
%
%
%
%
Therefore, this measurement dismisses the possibility that the IYR of $^{98}$Y is an outlier caused by false state ordering.

It is commonly accepted that FFs emerge from fission carrying several units of angular momentum, around 6-7 $\hbar$ for the spontaneous fission of $^{252}$Cf \cite{Wilhelmy,wilsonAngularMomentumGeneration2021,GaoAM,RakAM_tof,RakAM_piicr}.
Most of this angular momentum is removed by the $\gamma$-ray emission that follows the prompt neutron evaporation, as these neutrons are commonly assumed to not carry a significant amount of angular momentum.

Typically, each nucleus emits 3-4 photons before reaching one of the isomeric states. Due to selection rules, low multipolarity transitions are favored during the de-excitation process, with each photon carrying 1-2 $\hbar$. This qualitatively explains why the probability of populating a higher-spin state is larger for almost all FPs. This is however not true for $^{98}$Y.

Thanks to the use of two high-resolution methods in the same experimental setup, we can clearly observe that $^{98}$Y is the only yttrium isotope with an anomalously low IYR. Conversely, both $^{99}$Nb and $^{100}$Nb, \textit{i.e.} the transitional niobium isotopes at $N=60$, show no such behaviour.

Figure~\ref{fig:IYR_bib} shows the measured data together with literature data from different fissioning systems. 
It shows that this is not the first time such a low value for the IYR of $^{98}$Y has been observed. 
The IYR values reported in Refs. \cite{Tipnis,Bail,Martin} for various fissioning systems, show as well that the population of the low-spin state of $^{98}$Y was strongly favored. Note that the reason for the strong disagreement on the IYR of $^{99}$Nb in Figure~\ref{fig:IYR_bib} is not clearly identified, but might be related to an outdated decay scheme being used in the $\gamma$-spectroscopy measurements.

Most of the $^{98}$Y nuclei produced in fission actually come from neutron evaporation of the neutron-rich precursors, with significant contributions typically up to N+4 ($^{102}$Y). 
Both the prolate shape and the mean squared charge radii of the neutron-rich precursors of $^{98}$Y are more similar to those of $^{98m}$Y (see Fig.~\ref{fig:IYR}).
It is therefore again surprising that the excited $^{98}$Y*, before the $\gamma$-ray de-excitation, produced by the neutron emission of prolate nuclei most likely proceeds towards a smaller oblate shape.

What we present as observations of IYR results from an ensemble of highly-excited initial states of fragments that de-excite to the potential minima, i.e., the isomeric states of a nucleus. If it is possible to define a shape for excited fragments, it could be seen as a superposition of many states, that might, on average, even be
spherical. Thus, this poses the question about the de-excitation path taken by the fragments on a potential energy surface towards isomeric
states.

Finally, this result suggests that the IYR could be used to investigate other nuclei where shape coexistence is observed, such as the isomeric states of $^{79}$Zn.  
Additionally, anomalously low IYR could also be an indicator for identifying possible shape coexistence in other nuclei.




\section*{Acknowledgements}
 We acknowledge the staff of the Accelerator Laboratory of University of Jyväskylä (JYFL-ACCLAB) for providing a stable online beam.
 This work was supported by the Swedish Research Council
 (Ref. No. 2020-04238) and has received funding from the Euratom research and training program 2014-2018 under grant agreement No. 847594 (ARIEL).
 We acknowledge support from the Research Council of Finland under grant numbers 295207, 327629, and 354968, the European Union’s Horizon 2020 Research and Innovation Programme under Grant Agreement No's 771036 (ERC CoG MAIDEN) and 861198-LISA-H2020-MSCA-ITN-2019.

\bibliographystyle{elsarticle-num} 
\bibliography{bibliography}



\end{document}